\documentclass[[letter paper,14pt]{extarticle}
\usepackage[utf8]{inputenc} 
\usepackage{amsmath} 
\usepackage{amssymb} 
\usepackage{graphicx} 
\usepackage{float} 
\usepackage[inner=0.5in, outer=0.75in, top=0.5in, bottom=0.5in]{geometry}
\usepackage{natbib}
\usepackage{caption}
\usepackage{subcaption}
\usepackage[shortlabels]{enumitem}
\usepackage{algorithm}
\usepackage{algpseudocode}
\usepackage{multirow}

\title{Geometric Model Selection for Latent Space Network Models: Hypothesis Testing via Multidimensional Scaling and Resampling Techniques}
\author{Jieyun Wang$^{1}$, Anna L. Smith$^{1}$  \\
        \small $^{1}$University of Kentucky, Dr. Bing Zhang Department of Statistics 
}
\date{}

\begin{document}
\maketitle
\begin{abstract}
Latent space models assume that network ties are more likely between nodes that are closer together in an underlying latent space.
Euclidean space is a popular choice for the underlying geometry, but hyperbolic geometry can mimic more realistic patterns of ties in complex networks. 
To identify the underlying geometry, past research has applied non-Euclidean extensions of multidimensional scaling (MDS) to the observed geodesic distances: the shortest path lengths between nodes.
The difference in stress, a standard goodness-of-fit metric for MDS, across the geometries is then used to select a latent geometry with superior model fit (lower stress).
The effectiveness of this method is assessed through simulations of latent space networks in Euclidean and hyperbolic geometries.
To better account for uncertainty, we extend permutation-based hypothesis tests for MDS to the latent network setting.
However, these tests do not incorporate any network structure.
We propose a parametric bootstrap distribution of networks, conditioned on observed geodesic distances and the Gaussian Latent Position Model (GLPM).
Our method extends the Davidson-MacKinnon J-test to latent space network models with differing latent geometries.
We pay particular attention to large and sparse networks, and both the permutation test and the bootstrapping methods show an improvement in detecting the underlying geometry.\\

\end{abstract}\hspace{10pt}


\section{Introduction}

Networks are often modeled using latent space models where each node is assigned a position in a low-dimensional latent space (\cite{hoff_latent_2002}; \cite{krioukov_hyperbolic_2010}; \cite{sweet_latent_2020}; \cite{sosa_latent_2021}). The probability of having an edge between two nodes is proportional to the distance in the underlying space, so that pairs of nodes that are further apart are less likely to be connected. Euclidean space has been a popular default choice for the underlying space since the model was introduced by \cite{hoff_latent_2002}. Yet there is a growing interest in alternative geometries, such as hyperbolic space.

\cite{krioukov_hyperbolic_2010} argues that hyperbolic geometry is effective for complex networks: First, hyperbolic spaces expand faster than Euclidean spaces.  Hyperbolic spaces expand exponentially, while Euclidean spaces expand polynomially. Many real-world networks tend to be treelike, and the structure of trees is similar to hyperbolic geometry in the sense that the number of nodes grows exponentially with the depth. Second, common network features, such as a skewed degree distribution and high amounts of clustering, are naturally generated from the basic properties of hyperbolic geometry. Furthermore, \cite{smith_geometry_2019} shows that modeling networks in hyperbolic space is promising, since it results in networks with higher levels of degree centrality, betweenness, and closeness without dramatic losses in clustering or average path length.

Current approaches to determining which geometry underlies a particular observed network include: (i) comparing the fit of multi-dimensional scaling (MDS) of geodesic network distances across different geometries, as in \cite{papamichalis_latent_2022}; (ii) borrowing tools from metric spaces, such as $\delta$-hyperbolicity (\cite{kennedy_hyperbolicity_2013}; \cite{narayan_large-scale_2011}) or Ollivier-Ricci curvature (\cite{van_der_hoorn_ollivier-ricci_2021}); and (iii) curvature estimation, via a clique-based latent distance estimator (\cite{lubold_identifying_2022}; \cite{wilkins-reeves_asymptotically_2024}).

An MDS-based approach is appealing since it is metric-agnostic and requires minimal computational effort. We assess the effectiveness of the method proposed in \cite{papamichalis_latent_2022} for geometry selection through simulations of latent space networks in hyperbolic and Euclidean geometry. To make such geometry selection statistically principled, we further expand the approach to appropriately account for uncertainty. The standard way of accounting for statistical uncertainty in MDS is through permutation tests (\cite{mair_goodness--fit_2016} as well as \cite{mair_more_2022}) or bootstrap-based methods (\cite{mair_more_2022}; \cite{jacoby_bootstrap_2014}). However, these approaches are likely overly conservative, as they do not account for the type of structures we expect in network data. Our proposed method forms bootstrap distributions of network data, conditional on the observed shortest paths.  Our approach builds on the success of using bootstrap distributions of networks to assess network structure \cite{levin_bootstrapping_2021}.

In Section 2, we review MDS in hyperbolic and Euclidean spaces, show how comparisons of standard goodness-of-fit measures, stress and strain, can guide geometry choice, and discuss how permutation tests for quantifying MDS uncertainty can be adapted to the network setting. In Section 3, we introduce methods for quantifying uncertainty for MDS goodness-of-fit metrics, reviewing the permutation test and extending the bootstrap $J$-test to the network setting. In Section 4, we describe the latent space models in both hyperbolic and Euclidean space and demonstrate the efficiency and accuracy of the methods through simulations. Section 5 demonstrates the methods on several real-world datasets.

\section{Multi-dimensional scaling in the network setting}

Without loss of generality, we assume a network of size $N$ consists of a set of $N$ nodes $\mathcal{N}=\{n_1,n_2,\dots,n_N\}$. We refer to this network by an adjacency matrix $\mathbf{Y}=(Y_{ij})_{1\le i,j\le N}$. We focus on undirected, unweighted networks without self-loops, that is for $i,j=1,2,\dots N$
\begin{align}
    Y_{ij}&=Y_{ji}=\begin{cases}
        1 &\text{if there is an edge between $n_i$ and $n_j$, }i\neq j\\
        0 &\text{otherwise}
    \end{cases}\\
    Y_{ii}&=0
\end{align}
Moreover, only connected networks are considered in this study; that is, all the elements $\delta_{ij}$ in the geodesic distance matrix $\mathbf{D}(\mathbf{Y}) = (\delta_{ij})_{1\le i,j\le N}$ are non-negative finite numbers, where $\delta_{ij}$ is the shortest path length between $n_i$ and $n_j$.

Classical MDS (\cite[chap 12]{borg_modern_2007}), also known as principal coordinate analysis, uses observed dissimilarities to embed data in a lower-dimensional metric space.
Given a dissimilarity matrix $\boldsymbol{\Delta}$, the goal is to find a coordinate matrix $\mathbf{X}_{\mathbb M}$ in a low-dimensional manifold $\mathbb M$ that minimizes a target function which measures stress or strain. 
Stress serves as a goodness-of-fit measure for MDS. Although multiple definitions of stress have been proposed, we use the definition given by \cite{keller-ressel_hydra_2019}, 
$$S_{\mathbb{M}}(\mathbf{Y}) = \sqrt{\sum_{i,j}(\delta_{ij}-\hat{d}^{\mathbb M}_{ij})^2}$$
where $\mathbf{Y}$ is the adjacency matrix of the observed network, $\delta_{ij}$ is the geodesic distance or shortest path lengths between nodes $i$ and $j$, which serves as a measure of dissimilarity in this study, and $\hat{d}^{\mathbb M}_{ij}$ is the distance between points $x_i$ and $x_j$ calculated in manifold $\mathbb M$ based on the coordinate matrix $\mathbf{X}_{\mathbb M}$ recovered by MDS.

In the following subsections, we briefly review how stress has been used to identify underlying latent network geometry and whether methods for quantifying uncertainty associated with MDS can be directly translated to the network setting.

\subsection{Observed Stress Difference}

In \cite{papamichalis_latent_2022}, they compare the quality of MDS across different underlying geometries.
The stress from classical and hyperbolic MDS are compared directly. The geodesic distance matrix is used as the dissimilarity matrix and if $S_{\mathbb H^2}(\mathbf{Y})-S_{\mathbb R^2}(\mathbf{Y})<0 $, then the hyperbolic space is preferred as the underlying latent space. 

\subsection{Permutation Tests}
In the literature, both permutation tests and bootstrapping have been used to measure uncertainty associated with MDS goodness of fit metrics \citep{mair_goodness--fit_2016, mair_more_2022, farine_permutation_2022}.

Following \cite{mair_goodness--fit_2016}, a permutation test for goodness of fit for MDS specifies null and alternative hypotheses as
\begin{description}
    \item [$\mathbf{H_0}$ :] Stress/configuration are obtained from a random permutation of dissimilarities.
    \item [$\mathbf{H_A}$ :] Stress/configuration are obtained from something other than a random permutation of dissimilarities.
\end{description}

This null hypothesis is weakly informative and states that dissimilarities are exchangeable and that there is no structure beyond random chance. Translating these hypotheses to the comparison between MDS approaches and the latent space network setting, the null and alternative hypotheses are given as
\begin{description}
    \item[$\mathbf{H_0}$ :] Difference between hyperbolic and euclidean stresses/configurations is obtained as the difference between stress/configurations from a random permutation.
    \item[$\mathbf{H_A}$ :] Difference between hyperbolic and euclidean stresses/configurations is obtained from something other than a random permutation.
\end{description}
\cite{mair_goodness--fit_2016} provide two scenarios for setting up a permutation scheme:
\begin{description}
    \item  [\textbf{S1} :] In the case of directly observed dissimilarities, the elements of  $\boldsymbol{\Delta}$ can be permuted.
    \item  [\textbf{S2} :] For derived dissimilarities, they proposed a strategy for systematic column-wise permutations of the raw data.
\end{description}

\begin{figure}[h!]
    \centering
    \includegraphics[width=0.7\linewidth]{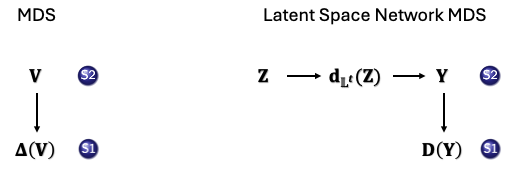}
    \caption{Permutation schemes for classical MDS (left) and in a latent space network setting (right).  On the right, the horizontal arrows represent the network generative process, where $\boldsymbol{Z}$ is a matrix of true underlying latent positions.}
    \label{fig:enter-label}
\end{figure}

Both schemes could be implemented in the latent space network setting, which is explained pictorially in Figure \ref{fig:enter-label}. Under the first scheme, the geodesic distance matrix is permuted directly, which would violate properties of the distance metric, such as the triangle inequality, and further weaken the null hypothesis. Instead, permuting the adjacency matrix $\mathbf{Y}$ allows us to preserve properties of the geodesic distance matrix; permutation of $\mathbf{Y}$ generates a new adjacency matrix,$\Tilde{\mathbf{Y}}$ which is then used to compute a corresponding geodesic distance matrix $\mathbf{D}(\Tilde{\mathbf{Y}})$. Since we only consider undirected networks, it is sufficient to permute only the elements within the upper triangle of the adjacency matrix $\mathbf{Y}$ which preserves symmetry in the new adjacency matrix $\Tilde{\mathbf{Y}}$. This method is described in Algorithm \ref{alg:permutation}.
\begin{algorithm}[h!]
	\caption{\textbf{Permutation Test: Density-preserving permuted networks}} 
	\begin{algorithmic}[1]
		\For {$i=1,2,\ldots,P$}
			\State Permute $\mathbf{Y}$ and get the new permuted adjacency matrix $\Tilde{\mathbf{Y}}_i$
            \If{New network with adjacency matrix $\Tilde{\mathbf{Y}}_i$ is connected} 
                \State Apply classical and hyperbolic MDS on $\mathbf{D}(\Tilde{\mathbf{Y}}_i)$
                \State Compute and store stress difference $S_{\mathbb H^2}(\Tilde{\mathbf{Y}}_i)-S_{\mathbb R^2}(\Tilde{\mathbf{Y}}_i)$
            \Else
                \State Discard the network and go to next iteration
            \EndIf
		\EndFor
        \State Calculate $P(S_{\mathbb H^2}(\Tilde{\mathbf{Y}})-S_{\mathbb R^2}(\Tilde{\mathbf{Y}})\le S_{\mathbb H^2}(\mathbf{Y})-S_{\mathbb R^2}(\mathbf{Y}))$.
        \If{$P(S_{\mathbb H^2}(\Tilde{\mathbf{Y}})-S_{\mathbb R^2}(\Tilde{\mathbf{Y}})\le S_{\mathbb H^2}(\mathbf{Y})-S_{\mathbb R^2}(\mathbf{Y}))<0.05$}
            \State Reject the null and hyperbolic geometry is preferred
        \Else
            \State Fail to reject the null and Euclidean geometry is preferred
        \EndIf
	\end{algorithmic} 
    \label{alg:permutation}
\end{algorithm}

\section{Bootstrapped networks for MDS}

The permuted $\Tilde{\mathbf{Y}}$ form a nonparametric bootstrap distribution of networks. While preserving edge density, the nonparametric bootstrap distribution will fail to preserve other features of the underlying latent network geometry.  For example, if two nodes have true latent positions which are close, they should be tied frequently in our bootstrap distribution.  Rather than permuting the edges at random, we consider whether MDS provides information about the underlying latent distances between nodes.

\cite{levin_bootstrapping_2021} provide two methods to bootstrap networks that leverage estimated underlying latent structure. The first method is designed to generate bootstrap replicates of U-statistics, which does not apply to our case. For the second method, they generate bootstrap replicates of whole networks and then evaluate network quantities. They work with the random dot product graph (RDPG) in which the latent positions $X_i$ are i.i.d. with inner product distribution $F$. In order to obtain bootstrapping replicates, they apply adjacency spectral embedding (ASE) on the adjacency matrix $A$, and get estimated latent positions $\Hat{X}_i$. Then, let $\Hat{F}_n$ denote the empirical distribution of $\Hat{X}_i$ and bootstrap samples, $X_i^*$, are drawn i.i.d. from $\Hat{F}_n$. Bootstrap samples generated in this way have adjacency matrices similar to the observed one.

In our setting, we apply MDS to the geodesic distance matrix, $\mathbf{D}(\mathbf{Y})$, and get a coordinate matrix, $X_{\mathbb{M}}$, recovered by MDS.  It is tempting to use $X_{\mathbb{M}}$ in place of the ASE-estimated latent positions in \citep{levin_bootstrapping_2021}'s approach and draw a bootstrap replicate network based on the empirical distribution of $X_{\mathbb{M}}$. However, this matrix does not contain estimates of the true latent positions, $\boldsymbol{Z}$. Recall that the latent space network model and the RDPG assume that the likelihood of \textit{a tie} between a pair of nodes is proportional to their latent distance.  However, the pairwise distances induced by $X_{\mathbb{M}}$ correspond to \textit{geodesic distances} on $Y$, and not tie presence directly. To generate bootstrap network samples with a geodesic distance matrix that have a distribution similar to the observed network, we will associate the geodesic distance with the underlying latent distance between two nodes. In this new bootstrapping method, we estimate the distribution of the distance between two nodes conditioned on their observed geodesic distance and generate the corresponding distance matrix. Then, generate a new bootstrap adjacency matrix.

\subsection{Geodesic distances given latent positions}
\cite{fronczak_average_2004} discuss an analytic solution for the average path length in a random graph, within which they provide a method to find the probability that the geodesic distance between two nodes is exactly $k$. \cite{rastelli_properties_2016} uses the same reasoning and finds an explicit expression under the Gaussian latent position model(Gaussian LPM, GLPM), which is given by:
    \begin{align}\label{eq:GLPM}
    \begin{split}
    \mathbf{z}_i & \in \mathbb R^d\\
    \mathbf{z}_i &\overset{iid}{\sim} \text{MVN}_2(\mathbf{0}, \gamma\mathbf{I}_2)\\
    Y_{ij}|\mathbf{z}_i,\mathbf{z}_j &\sim \text{Bernoulli}(p_{ij}), i\neq j\\
    p_{ij}& = \tau \exp\left\{-\frac{\left[d_{\mathbb R^d}(\mathbf{z}_i,\mathbf{z}_j)\right]^2}{2\phi}\right\}    
    \end{split}
    \end{align}
where $d$ is the dimensional of the underlying Euclidean space and in this study, we set $d=2$. $\gamma>0$ and is a parameter for the multivariate normal distribution, $\tau \in [0,1]$ and $\phi>0$, where $\tau$ controls the sparsity of the network and $\phi$ relates the probability of having an edge between two nodes with their distances in the latent space.

To summarize this method: First, define $p_{k}(\mathbf{z}_i,\mathbf{z}_j)$ as the probability of at least one walk of length $k$ between nodes $i$ and $j$, where $\mathbf{z}_i$ and $\mathbf{z}_j$ are their latent positions respectively. A walk is called a path only when all the nodes it passes through are distinct, while a walk does not require all its nodes to be distinct. In other words, $p_{k}(\mathbf{z}_i,\mathbf{z}_j)$ could also be thought of as the probability that the geodesic distance between these two nodes is not larger than $k$. Then, define $\ell_k(\mathbf{z}_i,\mathbf{z}_j)$ as the probability that the geodesic distance between nodes $i$ and $j$ is exactly $k$. Given the definition of $p_{k}(\mathbf{z}_i,\mathbf{z}_j)$, we have:
\begin{equation}
\ell_k(\mathbf{z}_i,\mathbf{z}_j) =p_k(\mathbf{z}_i,\mathbf{z}_j)-p_{k-1}(\mathbf{z}_i,\mathbf{z}_j)
\end{equation}
We further define $\xi_{k}(\mathbf{z}_i,\mathbf{z}_j)$ to be the probability that there exits a walk of length $k$ from node $i$ to node $j$ that passes through nodes $\{n_i,n_{A_m^1},\dots,n_{A_m^{k-1}},n_j\}$, and we refer to this walk of $k+1$ nodes as event $A_m$. By the definition of a walk, we do not require $n_{A_m^1},\dots,n_{A_m^{k-1}}$ to be distinct. The number of such events is about $N^{k-1}$. Let $A =  \cup_{m=1}^{N^{k-1}}A_m$, the event that at least one walk of length $k$ exists between nodes $i$ and $j$. By Lemma 1 of \cite{fronczak_average_2004}, we have:
\begin{equation}
p_k(\mathbf{z}_i,\mathbf{z}_j) \approx 1-\exp\{-N^{k-1}\xi_k(\mathbf{z}_i,\mathbf{z}_j)\}    
\end{equation}
Note that the assumption for that lemma is that all events $A_1, A_2,\dots$ need to be mutually independent. However, a same shorter walk could appear in multiple events; then, there is a correlation between these events. As \cite{fronczak_average_2004} points out, the proportion of such correlations is negligible for short walks ($k \ll N$).\\
Moreover, \cite{rastelli_properties_2016} has shown that the quantities above could be estimated using the following explicit expression: 

\begin{equation}
\begin{cases}
    h_{r+1} =   h_r\alpha_r^{-d}(2\pi\phi)^{\frac{d}{2}}f_d\left(\mathbf{z}_i;\mathbf{0},\frac{\omega_r+\gamma}{\alpha^2_r}\right)\\
    \alpha_{r+1}= \frac{\alpha_r \gamma}{\omega_r+\gamma}\\
    \omega_{r+1} = \frac{\omega_r\phi+\omega_r\gamma+\gamma\phi}{\omega_r+\gamma}
\end{cases}\text{, where}
\begin{cases}
    h_1 = \tau(2\pi\phi)^{\frac{d}{2}}\\
    \alpha_1 = 1\\
    \omega_1 = \phi
\end{cases}
\end{equation}
\begin{align}
\xi_k(\mathbf{z}_i,\mathbf{z}_j) &= h_kf_d\left(\mathbf{z}_j-\alpha_k\mathbf{z}_i;\mathbf{0},\omega_k\right)\text{, for }k=1,2,\dots,N-1\\
\begin{split}
\ell_k(\mathbf{z}_i,\mathbf{z}_j) &=p_k(\mathbf{z}_i,\mathbf{z}_j)-p_{k-1}(\mathbf{z}_i,\mathbf{z}_j)\\
&=\exp\{-N^{k-2}\xi_{k-1}(\mathbf{z}_i,\mathbf{z}_j)\}-\exp\{-N^{k-1}\xi_k(\mathbf{z}_i,\mathbf{z}_j)\}    
\end{split}
\end{align}

As suggested in \cite{rastelli_properties_2016}, for an observed network, assuming $\gamma=1$, $\phi$ and $\tau$ could be found via an ad-hoc method that matches the observed and theoretical values of average degree and the clustering coefficient through equation \eqref{eq:kbar} and \eqref{eq:cc}.

\subsection{The conditional latent distance distribution}
By definition, $\ell_k(\mathbf{z}_i,\mathbf{z}_j)$ is a conditional probability, that is,
\begin{equation}\label{eq:ell}
\ell_k(\mathbf{z}_i,\mathbf{z}_j) = P(\delta_{ij}=k|\mathbf{z}_i,\mathbf{z}_j).
\end{equation}
Based on the Gaussian LPM \eqref{eq:GLPM}, we know that the probability of having an edge depends only on $d_{ij} = d_{\mathbb R^d}(\mathbf{z}_i,\mathbf{z}_j)$. Thus, we could also rewrite equation \eqref{eq:ell} as
\begin{equation}
\ell_k(\mathbf{z}_i,\mathbf{z}_j) = P(\delta_{ij}=k|\mathbf{z}_i,\mathbf{z}_j)=P(\delta_{ij}=k|d_{ij}).    
\end{equation}
Then, the distribution of the latent distance between nodes given their geodesic distance could be found through: 
\begin{equation}\label{eq:M2}
    P(d_{ij}|\delta_{ij}=k) \propto P(\delta_{ij}=k|d_{ij})P(d_{ij})
\end{equation}
For the marginal distribution of latent distances, consider that when $d=2$ and $\gamma=1$,  
\begin{align}
d_{ij}&=|\mathbf{z}_i-\mathbf{z}_j| = \sqrt{(z_{i1}-z_{j1})^2+(z_{i2}-z_{j2})^2}\\
\frac{d_{ij}}{\sqrt{2}}&=\sqrt{\left(\frac{z_{i1}-z_{j1}}{\sqrt{2}}\right)^2+\left(\frac{z_{i2}-z_{j2}}{\sqrt{2}}\right)^2}
\end{align}
Based on the Gaussian LPM \eqref{eq:GLPM}, $z_{i1},z_{j1},z_{i2}$, and $z_{j2}$ are independent and identically distributed, and follow $N(0, 1)$. Then $\frac{z_{i1}-z_{j1}}{\sqrt{2}}$ and $\frac{z_{i2}-z_{j2}}{\sqrt{2}}$ are independent and identically distributed and follow $N(0,1)$. By the definition of the Chi distribution, we know that $d_{ij}/\sqrt{2}$ follows the Chi distribution with degree of freedom $2$, and $P(d_{ij})$ could easily be obtained.

\begin{figure}[h!]
  \centering

  \begin{subfigure}{0.95\textwidth}
    \centering
    \includegraphics[width=\textwidth]{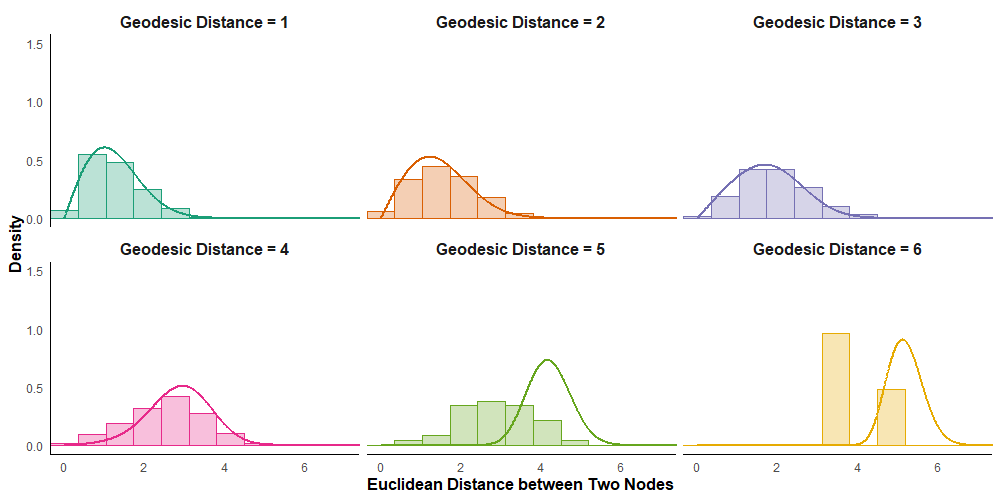}
    \caption{Network A, Density: 0.0988}
  \end{subfigure}

  \vspace{0.4cm}

  \begin{subfigure}{0.95\textwidth}
    \centering
    \includegraphics[width=\textwidth]{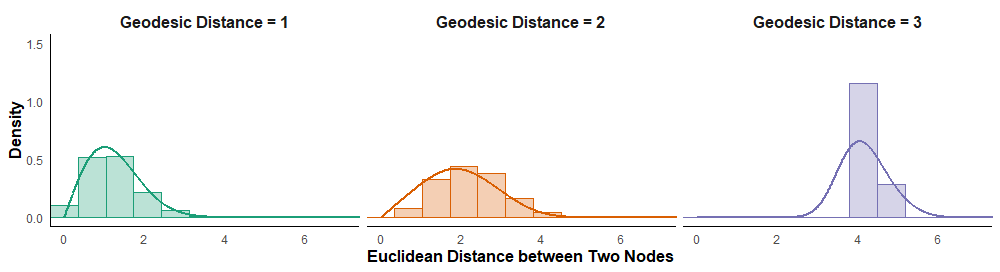}
    \caption{Network B, Density: 0.5176}
  \end{subfigure}

  \caption{Comparison between the theoretical conditional distribution of latent distances,shown by the curve, and the empirical histogram of observed latent distances between nodes. Both networks are of size 50 and generated using Gaussian LPM \eqref{eq:GLPM} with $\gamma=1$ and $\phi=2$, $\tau=0.2$ for Network A and $\tau=1$ for Network B.}
  \label{fig:conditional}
\end{figure}

The Figure \ref{fig:conditional} shows the comparison between the conditional latent distribution obtained through \eqref{eq:M2} and the histogram of latent distances between nodes given geodesic distance of two networks. Both networks are generated using Gaussian LPM, and the size is 50. In general, the theoretical distribution is broadly consistent with the empirical histograms, especially when geodesic distances are small. As the geodesic distance increases, the theoretical distributions become increasingly concentrated and shift toward larger latent distances. In contrast, the empirical histograms show more spread and include a substantial mass at smaller latent distances, suggesting that the theoretical model over-concentrates probability on larger distances and underrepresents shorter latent distances for higher geodesic levels. However, because these longer geodesic distances make up only a small proportion of the full distribution, this discrepancy is unlikely to substantially affect overall performance.

\subsection{Nonnested hypothesis tests}
The $J$-test is a classical way to compare two nonnested regression models. The core idea is that if model $M_2$ captures additional structure missing from model $M_1$, then $M_2$'s fitted values should add explanatory power. In Section 5 of \cite{davidson_bootstrap_2002}, they discuss the use of a parametric bootstrapped $J$ test, with hypotheses:
\begin{description}
    \item [$\mathbf{H_0}$ :] The competing model $M_2$ does not provide additional explanatory power beyond the current model $M_1$.
    \item [$\mathbf{H_A}$ :] The competing model $M_2$ provides additional explanatory power. 
\end{description}

To start with, they find the $J$ statistic for the observed data set, denoted by $\Hat{J}$. Then, they form a bootstrap distribution of the $J$ statistic. They fit the data into model $M_1$, generate parameter estimates, and use them in the bootstrap data generating process. Once they have $B$ bootstrap samples, $J$ statistics are calculated again for these samples using the same method as the observed data set and are denoted by $J_j^*, j=1,\dots,B$. Finally, the p-value for this test is given by
\begin{equation}
\Hat{p}^*(\hat{J}) = \frac{1}{B}\sum_{j=1}^B I(J^*_j\ge \Hat{J})  
\end{equation}

In the case of the latent space model, we adopt this bootstrapping method of the Davidson-MacKinnon $J$ test. Similar to the Davidson-MacKinnon $J$-test, we are comparing two nonnested models: GLPM as current model and a hyperbolic latent position model as competing model. The hypotheses are:

\begin{description}
    \item [$\mathbf{H_0}$ :] The hyperbolic model does not provide additional explanatory power beyond the euclidean model.
    \item [$\mathbf{H_A}$ ] The hyperbolic model provides additional explanatory power.
\end{description}

We first find the stress difference between classical and hyperbolic MDS. Then, just as they estimate model parameters to generate bootstrap data, we use the conditional distribution of the latent distance between nodes given geodesic distance under the GLPM to generate bootstrap networks under this model. Then, we calculate stress differences for these bootstrap networks and find the p-value. This method is described in Algorithm \ref{alg:bootstrap}.

\begin{algorithm}[h!]
	\caption{\textbf{Bootstrapped Networks: Conditional latent-distance sampling}}     
    \begin{algorithmic}[1]
        \For {$k=1,2,\dots K$}
        \State Compute and store $P(d_{ij}|\delta_{ij}=k)$
        \EndFor
		\For {$i=1,2,\ldots,B$}
			\State Generate distance matrix $\Tilde{\mathbf{d}}_{\mathbb R^2} = (\Tilde{d}_{ij})_{1\le i,j\le N}$. $\Tilde{d}_{ij}$ is drawn from $P(d_{ij}|\delta_{ij})$.
            \State Generate $\Tilde{\mathbf{Y}}_i$ based on $\Tilde{\mathbf{d}}_{\mathbb R^2}$ and GLPM 
            \If{New network with adjacency matrix $\Tilde{\mathbf{Y}}_i$ is connected} 
                \State Apply classical MDS and hydra on $\mathbf{D}(\Tilde{\mathbf{Y}}_i)$
                \State Compute and store stress difference $S_{\mathbb H^2}(\Tilde{\mathbf{Y}}_i)-S_{\mathbb R^2}(\Tilde{\mathbf{Y}}_i)$
            \Else
                \State Discard the network and go to next iteration
            \EndIf
		\EndFor
        \State Calculate $P(S_{\mathbb H^2}(\Tilde{\mathbf{Y}})-S_{\mathbb R^2}(\Tilde{\mathbf{Y}})\le S_{\mathbb H^2}(\mathbf{Y})-S_{\mathbb R^2}(\mathbf{Y}))$.
        \If{$P(S_{\mathbb H^2}(\Tilde{\mathbf{Y}})-S_{\mathbb R^2}(\Tilde{\mathbf{Y}})\le S_{\mathbb H^2}(\mathbf{Y})-S_{\mathbb R^2}(\mathbf{Y}))<0.05$}
            \State Reject the null and hyperbolic geometry is preferred
        \Else
            \State Fail to reject the null and Euclidean geometry is preferred
        \EndIf
	\end{algorithmic} 
    \label{alg:bootstrap}
\end{algorithm}

\section{Simulations}

We want to show the efficiency of the methods mentioned in the previous sections. To achieve this goal, we simulate networks in Euclidean and hyperbolic latent spaces, using the models described in the following sections. 

\subsection{Latent space models}

Modeling assumptions for the latent space models are as follows:
\begin{itemize}
    \item Each node $n_i$ has an unknown position $\mathbf{z}_i$ in $t$-dimensional latent space $\mathbb L^t$
    \item Latent positions $\mathbf{z}_i$'s are independent and follow some distributions in $\mathbb L^t$
    \item Edges are conditionally independent given latent positions and the probability of an edge between node $n_i$ and $n_j$ depends on a function $s_{\mathbb L^t}(\mathbf{z}_i,\mathbf{z}_j)$.
\end{itemize}

\subsubsection{Euclidean Geometry}\label{sec:glpm}
Most commonly mentioned latent space models for networks that use Euclidean geometry as underlying latent space is the "Distance Models" proposed in \cite{hoff_latent_2002}. Later, \cite{rastelli_properties_2016} proposed the Gaussian latent position model, in which a non-normalized Gaussian density is used for edge instead of logistic link function in Distance Models. We will use the Gaussian LPM \eqref{eq:GLPM} as the Euclidean latent space models. \cite{rastelli_properties_2016} has also shown that both expected average degree and clustering coefficient have explicit forms for Gaussian LPM, and are given by:
\begin{align}
\label{eq:kbar}
\Bar{k} &= (N-1)\tau\left\{\frac{\phi}{2\gamma+\phi}\right\}^{\frac{d}{2}}\\
\label{eq:cc}
C &= \tau\left(\frac{\gamma+\phi}{3\gamma+\phi}\right)^{\frac{d}{2}}
\end{align}

\subsubsection{Hyperbolic Geometry}\label{sec:hyp}
 The underlying geometry for latent space models is not limited to Euclidean geometry. \cite{krioukov_hyperbolic_2010} proposes a framework that assumes that the underlying geometry for complex networks is hyperbolic geometry. The hyperbolic latent space model is given by:
    \begin{align}
    \begin{split}
    \mathbf{z}_i &= (r_i, \theta_i)' \in \mathbb H^d\\
    r_i&\overset{iid}{\sim} \text{Uniform}(0, R)\\
    \theta_i &\overset{iid}{\sim} \text{Uniform}(0,2\pi)\\
    Y_{ij}|\mathbf{z}_i,\mathbf{z}_j &\sim \text{Bernoulli}(p_{ij}), i\neq j\\
    logit(p_{ij})& = R-d_{\mathbb H^d}(\mathbf{z}_i,\mathbf{z}_j)     
    \end{split}
    \end{align}
where $d$ is the dimensional of the underlying hyperbolic space and in this study, we set $d=2$. $R$ is the intrinsic radius, which controls how much latent space is being used. The latent positions are only sampled within disks of intrinsic radius $R$.

\subsection{Network measures}
The following network summary measures are used:
\begin{itemize}
    \item Expected Average Degree $\Bar{k}$, the expected number of connected nodes that each node has. 
    \item Density $den(\mathbf{Y})$: the proportion of all possible edges observed in a network.
    \item Clustering coefficient $C$: if $n_i$ and $n_j$ are connected, and if $n_j$ and $n_k$ are connected, the clustering coefficient is the probability that $n_i$ and $n_k$ are connected. 
\end{itemize}

\subsection{Simulation details}

For hyperbolic networks, based on \cite{krioukov_hyperbolic_2010}, if the network size is $N$, set $R = 2\log\left(\frac{8n}{\pi \bar k}\right)$, where $\bar{k}$ is the target average degree and in this simulation study, by varying $\Bar{k}$, we are able to generate networks for different densities. For GLPM, we choose $\gamma=1$ and $\phi=2$, and varying $\tau$ to get networks of various densities. 

To evaluate the efficiency of comparing raw stress difference, we tested networks of size 15 to 75 with densities ranging from 0 to 0.9. To compare all three methods, we first tested networks of size 15 to 45 with densities ranging from 0 to 0.9. Then, we focus on large and sparse networks whose sizes ranges from 50 to 200 and densities up to 0.2.

We apply classical MDS as implemented in the \texttt{stats} package in R.
For MDS in hyperbolic space, we rely on the \texttt{hydra} package in R \cite{keller-ressel_hydra_2019}. 

\subsection{Results}

\subsubsection{Comparing raw stress differences}
\begin{figure}[h!]
    \centering
    \includegraphics[width=\textwidth]{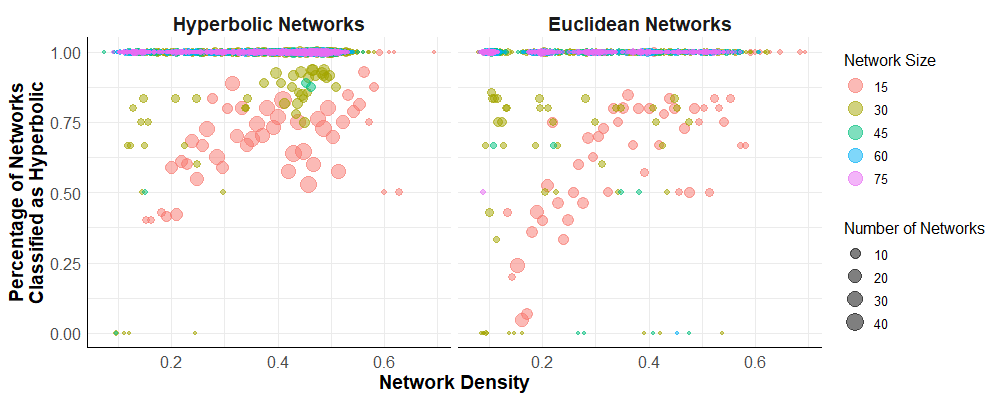}
    \caption{Comparing raw stress differences misclassifies almost all Euclidean networks as hyperbolic.}
    \label{fig: m0}
\end{figure}

As shown in Figure \ref{fig: m0}, we notice that as network size increases, hyperbolic networks tend to be correctly identified.  When network size is larger than 60, almost all hyperbolic networks are correctly categorized as hyperbolic. For smaller hyperbolic networks, the probability of correctly identifying hyperbolic geometry gets larger as the density of the network gets larger. However, this method doesn't work well for Euclidean networks, and larger and/or denser networks tends to be misclassified as hyperbolic network.
\subsubsection{Adding uncertainty from MDS}
We compare this approach to the above methods for accounting for uncertainty from MDS: (1) classical permutation tests for MDS and (2) bootstrapped networks from observed geodesic distances. We adopt the language of diagnostic tests. Let sensitivity indicate how well a method correctly identifies hyperbolic networks, and specificity indicate how well a method correctly identifies Euclidean networks.

First, we focus on smaller networks with a wide range of densities in Table \ref{tab:comp1}.

\begin{table}[h]
    \centering
    \resizebox{\textwidth}{!}{
    \begin{tabular}{c|c||cc|cc|cc}
        \multirow{2}{*}{\textbf{N}}&\multirow{2}{*}{\textbf{Density}} & \multicolumn{2}{c|}{$S_{\mathbb{H}^2}(\mathbf{Y}) - S_{\mathbb{R}^2}(\mathbf{Y})$} &  \multicolumn{2}{c|}{\textbf{Permutation Test}}  &  \multicolumn{2}{c}{\textbf{Bootstrapped Networks}}\\
        &&Sensitivity & Specificity  & Sensitivity & Specificity  & Sensitivity & Specificity  \\
        \hline \hline
        \multirow{3}{*}{15}
        &0-0.2 & 0.4516 & 0.7593 & 0.1936 & 0.9537 & 0.2273 & 1\\
         &0.2-0.4  & 0.6905 & 0.3908 & 0.184 & 0.937 & 0.163 & 0.9474 \\
         &0.4-1  & 0.714 & 0.2273 & 0.0828 & 0.9221 & 0.1037 & 0.9\\
         \hline
         \multirow{3}{*}{30}
         &0-0.2  & 0.8889 & 0.2158 & 0.3241 & 0.8633 & 0.2222 & 0.8861\\
         &0.2-0.4   & 0.9681 & 0.0677 & 0.4867 & 0.7961 & 0.3032 & 0.8469 \\
         &0.4-1  & 0.9457 & 0.0581 & 0.2403 & 0.8968 & 0.2327 & 0.8903\\
         \hline
         \multirow{3}{*}{45}
         &0-0.2  & 0.9823 & 0.0072 & 0.6372 & 0.8129 & 0.354 & 0.8208\\
         &0.2-0.4  & 1 & 0.0242 & 0.7247 & 0.6667 & 0.5899 & 0.7621 \\
         &0.4-1  & 0.9962 & 0.013 & 0.3277 & 0.8571 & 0.3089 & 0.8896\\
    \end{tabular}
    }
    \caption{Smaller Networks}
    \label{tab:comp1}
\end{table}

In general, for smaller networks, comparing raw stress differences identifies hyperbolic networks very well but, at the same time, tends to misclassify the majority of Euclidean networks especially as the networks get larger and/or denser. Both the permutation test and bootstrapped approach are more conservative and less likely to falsely categorize Euclidean networks as hyperbolic networks but misclassify the hyperbolic networks more than roughly half the time.

Then to mimic many real world networks, we focused on a particular group of networks - large and sparse networks. In this case, only networks of densities between 0 and 0.2 will be considered. 

\begin{table}[h]
    \centering
    \begin{tabular}{c||cc|cc|cc}
        \multirow{2}{*}{\textbf{N}} & \multicolumn{2}{c|}{$S_{\mathbb{H}^2}(\mathbf{Y}) - S_{\mathbb{R}^2}(\mathbf{Y})$} &  \multicolumn{2}{c|}{\textbf{Permutation Test}}  &  \multicolumn{2}{c}{\textbf{Bootstrapped Networks}} \\
        &Sensitivity & Specificity &Sensitivity & Specificity & Sensitivity & Specificity \\
        \hline \hline
        50 &0.9933  & 0.0333 & 0.6667 & 0.77333 & 0.4200 & 0.7750 \\
        \hline
        100 & 1 & 0 &  0.94& 0.6133 & 0.6133 & 0.6871 \\
        \hline
        150 & 1 & 0  &  0.98& 0.58 & 0.7333 & 0.6933\\
        \hline
        200 & 1 & 0 & 0.9867 & 0.7 & 0.84 & 0.7933\\
    \end{tabular}
    \caption{Large and Sparse Networks}
    \label{tab:comp2}
\end{table}

From Table \ref{tab:comp2}, when network size is 100 or larger, comparing raw stress differences categorizes all simulated networks, hyperbolic and Euclidean, as hyperbolic networks and cannot distinguish underlying geometry. Both the permutation test and bootstrapped based approach show an improvement in distinguishing between the two geometries for large and sparse network, which is more common in real-world datasets. The permutation test has better sensitivity, meaning that it is good at correctly identifying hyperbolic networks, while incorporating a bootstrapped network distribution results in better balance between sensitivity and specificity. 

\section{Real Data}
\subsection{Karate Club}
\begin{figure}[h!]
    \centering
    \begin{subfigure}[b]{0.45\textwidth}
        \centering
        \includegraphics[width=\textwidth]{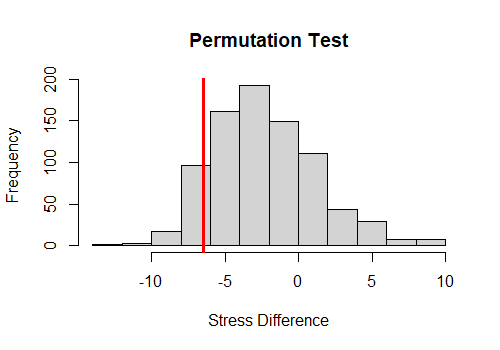}
        \caption{p-value: 0.1053}
    \end{subfigure}
    \begin{subfigure}[b]{0.45\textwidth}
        \centering
        \includegraphics[width=\textwidth]{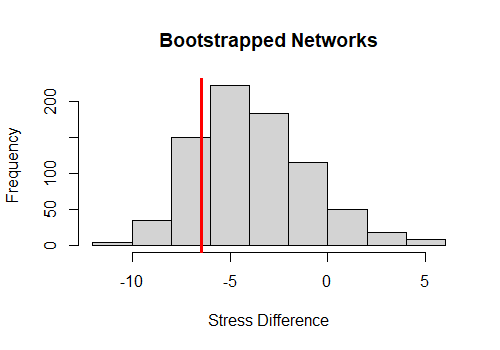}
        \caption{p-value: 0.1888}
    \end{subfigure}
    \caption{Stress differences across permuted (left panel) and bootstrapped (right panel) networks based on Zachary's Karate network.  All differences are calculated as $S_{\mathbb{H}^2} - S_{\mathbb{R}^2}$, with a negative difference indicating that hyperbolic space better describes the observed network geodesics. The vertical red lines correspond to the raw stress difference, without accounting for uncertainty.}
    \label{fig: karateynb}
\end{figure}

\cite{papamichalis_latent_2022} use Zachary’s Karate club network as an example. They compare the quality of MDS estimates when the manifold is assumed to be hyperbolic and Euclidean, and conclude that a model with a hyperbolic latent space better suits the data. Using the same package mentioned in this paper, \texttt{hydra}, we find $S_{\mathbb R^2}(\mathbf{Y}) = 24.65$ and $S_{\mathbb H^2}(\mathbf{Y})=18.20$. The stress from hyperbolic space is smaller, which would seem to indicate a higher quality of embedding and that hyperbolic geometry would be better to model the data. Accounting for uncertainty for MDS, we get p-values of 0.1053 from classical MDS permutation tests and 0.1888 from our bootstrap network distribution. This leads us to the conclusion that a Euclidean geometry is better.  In Figure \ref{fig: karateynb}, we show the distribution of stress differences across the permuted networks (panel a) and bootstrapped networks (panel b). 

\subsection{Other Datasets}
\label{sec:otherdata}
We have also applied these methods to four additional network datasets \citep{nepusz_fuzzy_2008, newman_finding_2006, lusseau_bottlenose_2003,kadushin_friendship_1995}.  A comparison of their network characteristics and the results from each of the three different methods is presented below. These networks are relatively sparse and range in size from 28 to 112 nodes.   

Across the four networks, three show agreement among all applicable methods on the underlying geometry, and the exception is the French Financial Elite Friendships network (ffe\_friend). Similar to Zachary's Karate network, the raw stress difference method favors a hyperbolic embedding. However, both methods accounting for uncertainty for MDS indicate that a Euclidean geometry provides the better fit. Overall, based on the permutation test and the bootstrapped network methods, UKfaculty and adjnoun are better modeled with hyperbolic geometry, whereas dolphins and ffe\_friend are better with Euclidean.

For the UKfaculty network, the bootstrapped network method is not applicable. Several nodes, for example node 73, have very low degree (connected to only two others), and those neighbors are also sparsely connected. This places such nodes far from the rest in latent space, lowers their probability of forming ties, and rarely yields a fully connected graph. Because the bootstrapped network method requires the resampled graph to be fully connected, it is difficult to implement for this dataset.

 \begin{figure*}
        \centering
        \begin{subfigure}[b]{0.475\textwidth}
            \centering
            \includegraphics[width=\textwidth]{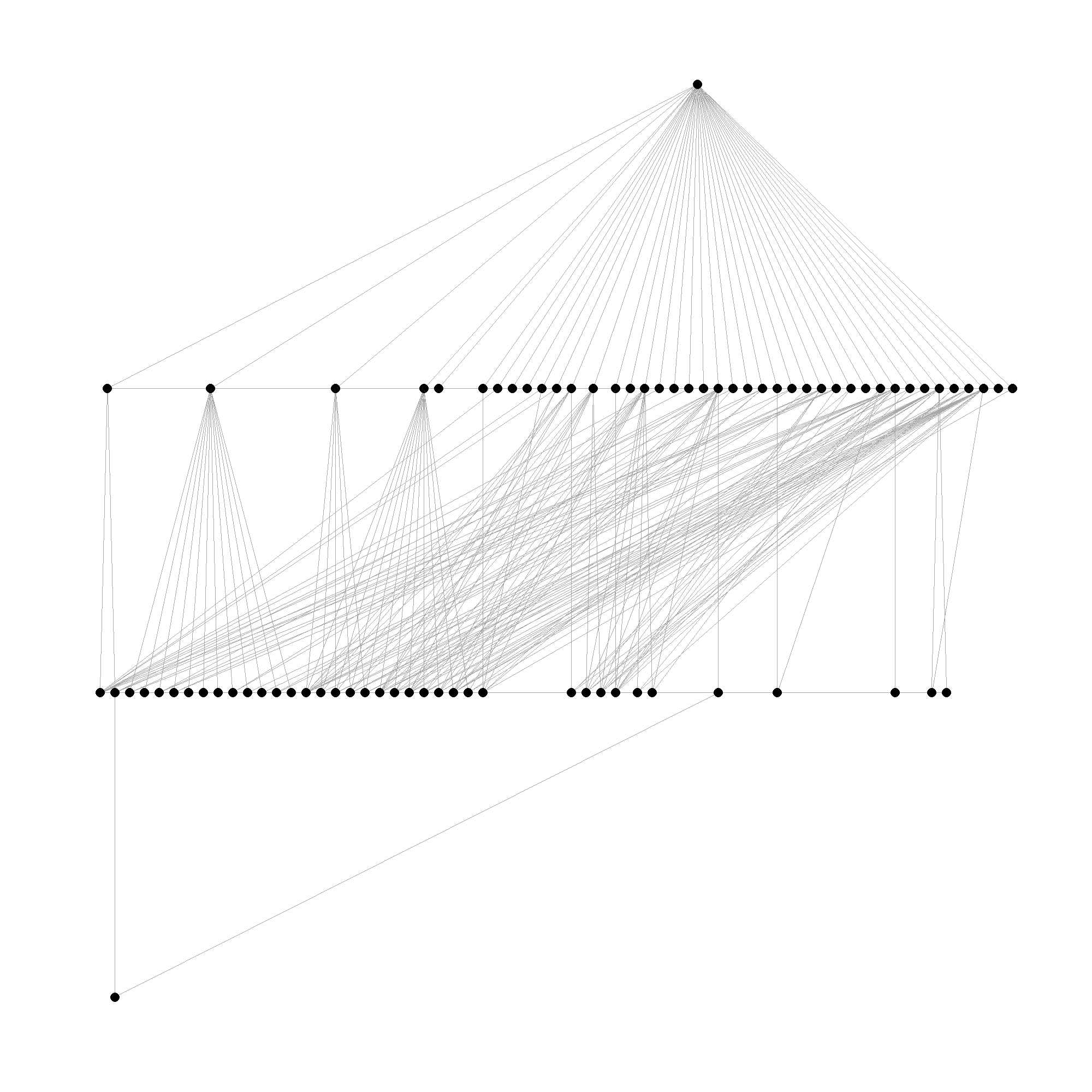}
            \caption[]%
            {{\small UKfaculty ($n=81$, dens.$=0.178$)}\\[-.05in]

            \centering
            \begin{tabular}{c|c|c}
                $S_{\mathbb{H}^2}-S_{\mathbb{R}^2}$ & Permutation & Bootstrap \\
                \hline
                -19.32 & p-value: 0 & - \\

                $\mathbb{H}^2$ & $\mathbb{H}^2$ & -
            \end{tabular}

            }
            \label{fig:mean and std of net14}
        \end{subfigure}
        \hfill
        \begin{subfigure}[b]{0.475\textwidth}  
            \centering 
            \includegraphics[width=\textwidth]{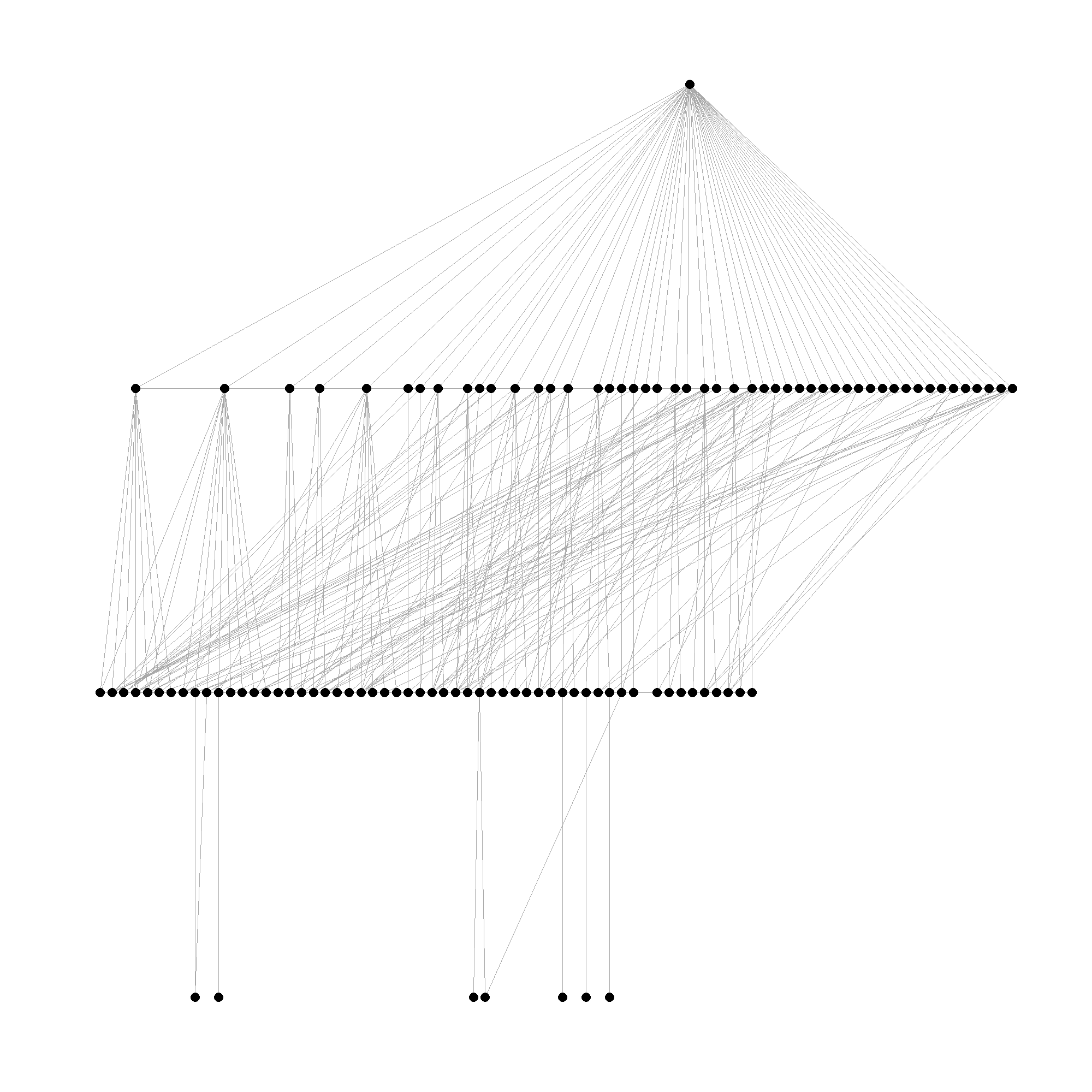}
            \caption[]%
            {{\small adjnoun ($n=112$, dens.$=0.068$)}\\[-.05in]
            
            \centering
            \begin{tabular}{c|c|c}
                $S_{\mathbb{H}^2}-S_{\mathbb{R}^2}$ & Permutation & Bootstrap \\
                \hline
                -62.61 & p-value: 0 & p-value: 0 \\
                $\mathbb{H}^2$ & $\mathbb{H}^2$ & $\mathbb{H}^2$
            \end{tabular}

            }    
            \label{fig:mean and std of net24}
        \end{subfigure}
        \vskip\baselineskip
        \begin{subfigure}[b]{0.475\textwidth}   
            \centering 
            \includegraphics[width=\textwidth]{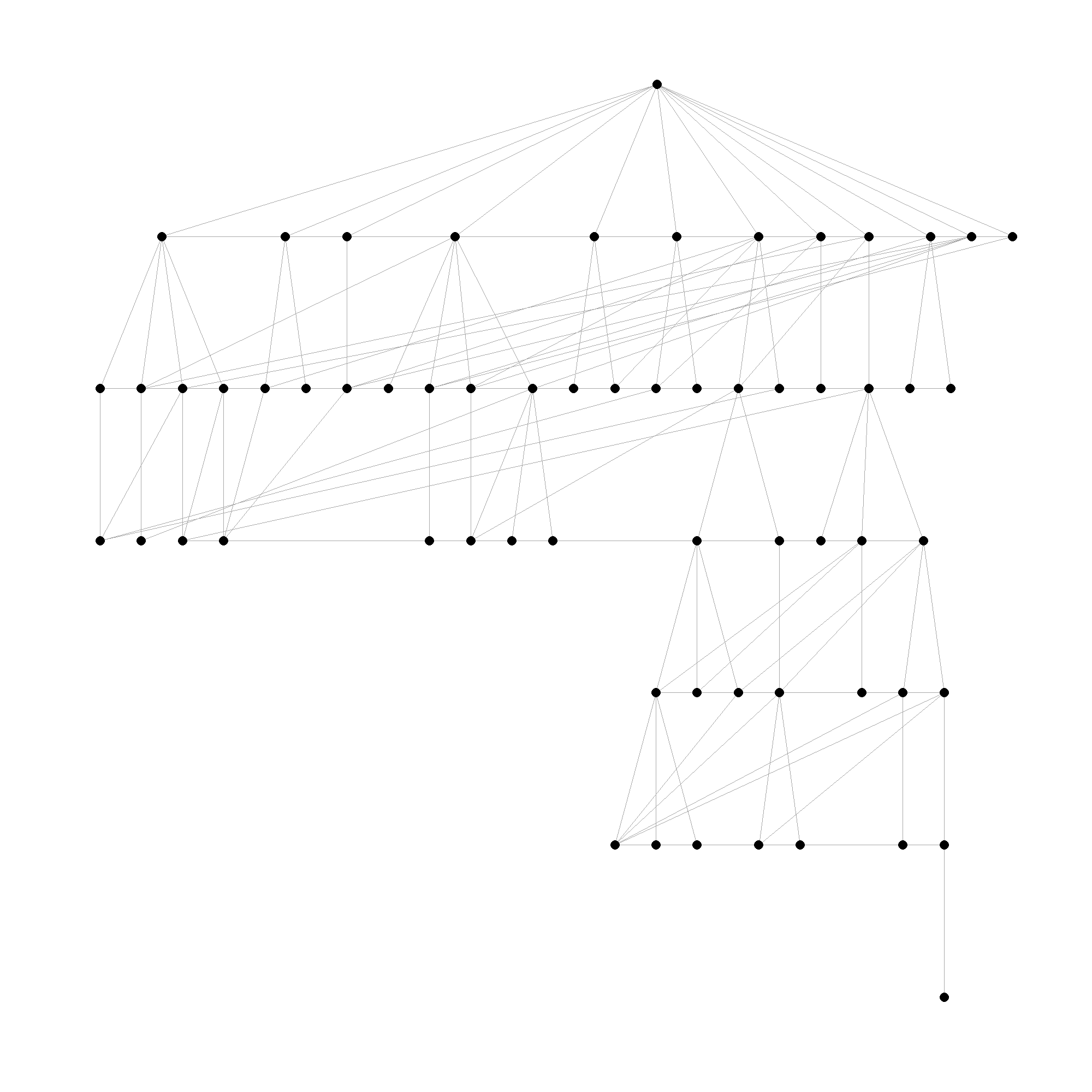}
            \caption[]%
            {{\small dolphins ($n=62$, dens.$=0.084$)}\\[-.05in]
            
            \centering
            \begin{tabular}{c|c|c}
                $S_{\mathbb{H}^2}-S_{\mathbb{R}^2}$ & Permutation & Bootstrap \\
                \hline
                11.51 & p-value: 0.99 & p-value: 0.99 \\

                $\mathbb{R}^2$ & $\mathbb{R}^2$ & $\mathbb{R}^2$
            \end{tabular}
            }    
            \label{fig:mean and std of net34}
        \end{subfigure}
        \hfill
        \begin{subfigure}[b]{0.475\textwidth}   
            \centering 
            \includegraphics[width=\textwidth]{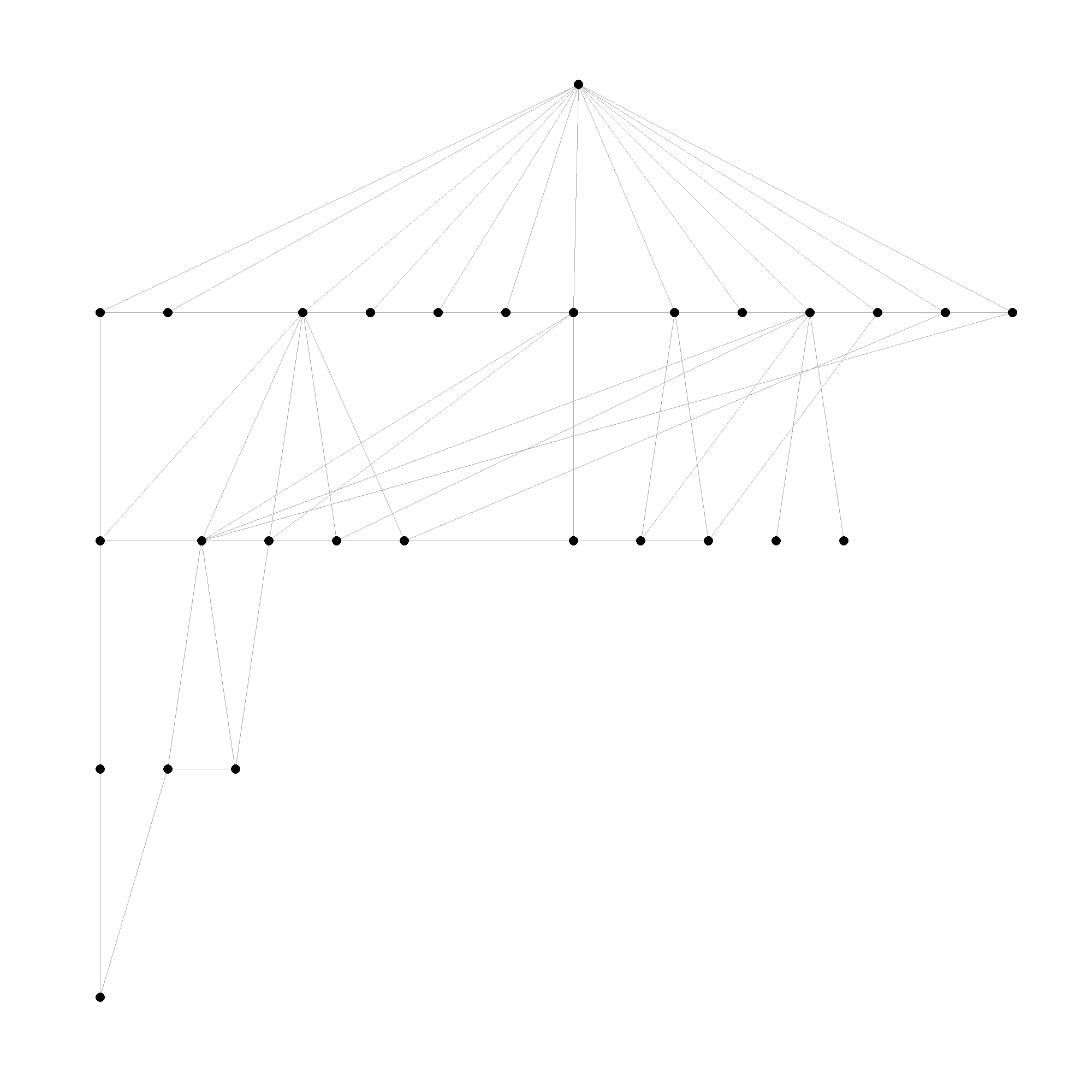}
            \caption[]%
            {{\small ffe\_friend ($n=28$, dens.$=0.175$)}\\[-.05in]
            
            \centering
            \begin{tabular}{c|c|c}
                $S_{\mathbb{H}^2}-S_{\mathbb{R}^2}$ & Permutation & Bootstrap \\
                \hline
                -3.60 & p-value: 0.20 & p-value: 0.38 \\

                $\mathbb{H}^2$ & $\mathbb{R}^2$ & $\mathbb{R}^2$
            \end{tabular}
            }    
            \label{fig:mean and std of net44}
        \end{subfigure}
        \caption{MDS-based classifications of a collection of real world networks.  All network plots are created using package \texttt{igraph}, which emphasizes tree-like behavior more common in hyperbolic networks.  Each network plot is annotated with its network size, $n$, edge density (dens.), and a table describing the results of the MDS-based classifications.} 
        \label{fig:mean and std of nets}
    \end{figure*}

\section{Discussion}

We studied the effectiveness of using observed stress difference from MDS in determining the underlying geometry for latent space models through a simulation study. We have demonstrated that when the stress difference is used alone, the geometry cannot be correctly identified.  In fact, comparing the stress difference directly nearly always misclassifies large Euclidean networks as hyperbolic.

Two new methods based on the stress difference from MDS were proposed that account for uncertainty with permutation tests and bootstrapped networks. By studying the same sets of simulated networks, we showed that both methods outperform the original simple stress difference, and work better on large and sparse networks which are commonly seen in real world.  However, both permutation and bootstrapping methods are computationally expensive since both classical MDS and hyperbolic MDS need to be applied to each of the permuted or bootstrapping networks. 

The advantage of the permutation test is that it retains the density of the observed network and doesn't rely on the latent space models, but the structure of the network is not considered when generating the permuted networks. On the other hand, the bootstrap method takes the structure of the observed networks into account: if two nodes are currently connected, they are likely closer in the latent space, and thus more likely to be connected in the bootstrapped networks. In this study, we used the Gaussian LPM as the Euclidean latent space model. However, for some networks, $\phi$ and $\tau$ could not be estimated using the ad-hoc method described in \citet{rastelli_properties_2016}, thus bootstrapping could not be implemented. Another difficulty in implementing the bootstrap method is that it sometimes produces networks that are not fully connected. As shown above for the UKfaculty network in Section \ref{sec:otherdata}, bootstrap resampling frequently yielded disconnected networks, consequently, the procedure could not be executed and no p-value could be computed. One potential solution is to remove isolated nodes, but this results in a bootstrapped networks that can have varying network sizes, which is undesirable. 

The bootstrap method could be extended to other network models, including more general Euclidean latent space models (i.e., that do not assume a Gaussian link) and latent space models in curved geometries. This requires deriving the model-based conditional distribution of the latent distance between nodes given their geodesic distance.  In cases where this can not be found theoretically, Monte Carlo estimation could be used. In the current study, we compare hyperbolic and Euclidean latent space models, using Euclidean geometry as the baseline and hyperbolic geometry in the competing model. A natural extension would be to adopt a spherical geometry as the baseline. Because spherical and hyperbolic spaces are more geometrically distinct than Euclidean and hyperbolic spaces, this choice may further improve the ability to identify networks with underlying hyperbolic structure.

\newpage
\bibliography{network}

@article{smith_geometry_2019,
	title = {The {Geometry} of {Continuous} {Latent} {Space} {Models} for {Network} {Data}},
	volume = {34},
	issn = {0883-4237},
	url = {https://projecteuclid.org/journals/statistical-science/volume-34/issue-3/The-Geometry-of-Continuous-Latent-Space-Models-for-Network-Data/10.1214/19-STS702.full},
	doi = {10.1214/19-STS702},
	number = {3},
	urldate = {2024-11-27},
	journal = {Statistical Science},
	author = {Smith, Anna L. and Asta, Dena M. and Calder, Catherine A.},
	month = aug,
	year = {2019},
}

@article{hoff_latent_2002,
	title = {Latent {Space} {Approaches} to {Social} {Network} {Analysis}},
	volume = {97},
	issn = {0162-1459, 1537-274X},
	url = {http://www.tandfonline.com/doi/abs/10.1198/016214502388618906},
	doi = {10.1198/016214502388618906},
	language = {en},
	number = {460},
	urldate = {2024-11-27},
	journal = {Journal of the American Statistical Association},
	author = {Hoff, Peter D and Raftery, Adrian E and Handcock, Mark S},
	month = dec,
	year = {2002},
	pages = {1090--1098},
}

@article{krioukov_hyperbolic_2010,
	title = {Hyperbolic {Geometry} of {Complex} {Networks}},
	volume = {82},
	issn = {1539-3755, 1550-2376},
	url = {http://arxiv.org/abs/1006.5169},
	doi = {10.1103/PhysRevE.82.036106},
	number = {3},
	urldate = {2024-11-27},
	journal = {Physical Review E},
	author = {Krioukov, Dmitri and Papadopoulos, Fragkiskos and Kitsak, Maksim and Vahdat, Amin and Boguna, Marian},
	month = sep,
	year = {2010},
	note = {arXiv:1006.5169 [cond-mat]},
	pages = {036106},
}

@misc{papamichalis_latent_2022,
	title = {Latent {Space} {Network} {Modelling} with {Hyperbolic} and {Spherical} {Geometries}},
	url = {http://arxiv.org/abs/2109.03343},
	doi = {10.48550/arXiv.2109.03343},
	urldate = {2024-11-27},
	publisher = {arXiv},
	author = {Papamichalis, Marios and Turnbull, Kathryn and Lunagomez, Simon and Airoldi, Edoardo},
	month = feb,
	year = {2022},
	note = {arXiv:2109.03343 [stat]},
}

@misc{keller-ressel_hydra_2019,
	title = {Hydra: {A} method for strain-minimizing hyperbolic embedding of network- and distance-based data},
	shorttitle = {Hydra},
	url = {http://arxiv.org/abs/1903.08977},
	doi = {10.48550/arXiv.1903.08977},
	urldate = {2024-11-27},
	publisher = {arXiv},
	author = {Keller-Ressel, Martin and Nargang, Stephanie},
	month = sep,
	year = {2019},
	note = {arXiv:1903.08977 [stat]},
}

@book{borg_modern_2007,
	address = {Dordrecht},
	edition = {2nd ed},
	series = {Springer {Series} in {Statistics}},
	title = {Modern {Multidimensional} {Scaling}: {Theory} and {Applications}},
	isbn = {978-0-387-25150-9},
	shorttitle = {Modern {Multidimensional} {Scaling}},
	language = {eng},
	publisher = {Springer-Verlag New York Inc},
	author = {Borg, Ingwer and Groenen, Patrick},
	year = {2007},
}

@article{farine_permutation_2022,
	title = {Permutation tests for hypothesis testing with animal social network data: {Problems} and potential solutions},
	volume = {13},
	issn = {2041-210X, 2041-210X},
	shorttitle = {Permutation tests for hypothesis testing with animal social network data},
	url = {https://besjournals.onlinelibrary.wiley.com/doi/10.1111/2041-210X.13741},
	doi = {10.1111/2041-210X.13741},
	language = {en},
	number = {1},
	urldate = {2024-12-02},
	journal = {Methods in Ecology and Evolution},
	author = {Farine, Damien R. and Carter, Gerald G.},
	month = jan,
	year = {2022},
	pages = {144--156},
}

@article{mair_goodness--fit_2016,
	title = {Goodness-of-{Fit} {Assessment} in {Multidimensional} {Scaling} and {Unfolding}},
	volume = {51},
	issn = {0027-3171},
	url = {https://www.tandfonline.com/doi/abs/10.1080/00273171.2016.1235966},
	doi = {10.1080/00273171.2016.1235966},
	number = {6},
	journal = {Multivariate Behavioral Research},
	author = {Mair, Patrick and Borg, Ingwer and Rusch, Thomas},
	month = nov,
	year = {2016},
	note = {Publisher: Routledge},
	pages = {772--789},
}

@misc{levin_bootstrapping_2021,
	title = {Bootstrapping {Networks} with {Latent} {Space} {Structure}},
	url = {http://arxiv.org/abs/1907.10821},
	doi = {10.48550/arXiv.1907.10821},
	urldate = {2024-12-02},
	publisher = {arXiv},
	author = {Levin, Keith and Levina, Elizaveta},
	month = oct,
	year = {2021},
	note = {arXiv:1907.10821 [math]},
}

@article{mair_more_2022,
	title = {More on {Multidimensional} {Scaling} and {Unfolding} in \textit{{R}} : \textbf{smacof} {Version} 2},
	volume = {102},
	issn = {1548-7660},
	shorttitle = {More on {Multidimensional} {Scaling} and {Unfolding} in \textit{{R}}},
	url = {https://www.jstatsoft.org/v102/i10/},
	doi = {10.18637/jss.v102.i10},
	language = {en},
	number = {10},
	urldate = {2024-12-04},
	journal = {Journal of Statistical Software},
	author = {Mair, Patrick and Groenen, Patrick J. F. and De Leeuw, Jan},
	year = {2022},
}

@article{rastelli_properties_2016,
	title = {Properties of latent variable network models},
	volume = {4},
	copyright = {https://www.cambridge.org/core/terms},
	issn = {2050-1242, 2050-1250},
	doi = {10.1017/nws.2016.23},
	language = {en},
	number = {4},
	urldate = {2024-12-04},
	journal = {Network Science},
	author = {Rastelli, Riccardo and Friel, Nial and Raftery, Adrian E.},
	month = dec,
	year = {2016},
	pages = {407--432}
}

@article{fronczak_average_2004,
	title = {Average path length in random networks},
	volume = {70},
	copyright = {http://link.aps.org/licenses/aps-default-license},
	issn = {1539-3755, 1550-2376},
	url = {https://link.aps.org/doi/10.1103/PhysRevE.70.056110},
	doi = {10.1103/PhysRevE.70.056110},
	language = {en},
	number = {5},
	urldate = {2025-01-23},
	journal = {Physical Review E},
	author = {Fronczak, Agata and Fronczak, Piotr and Hołyst, Janusz A.},
	month = nov,
	year = {2004},
	pages = {056110},
}

@article{davidson_bootstrap_2002,
	title = {Bootstrap {J} tests of nonnested linear regression models},
	volume = {109},
	copyright = {https://www.elsevier.com/tdm/userlicense/1.0/},
	issn = {03044076},
	url = {https://linkinghub.elsevier.com/retrieve/pii/S0304407601001464},
	doi = {10.1016/S0304-4076(01)00146-4},
	language = {en},
	number = {1},
	urldate = {2025-05-15},
	journal = {Journal of Econometrics},
	author = {Davidson, Russell and MacKinnon, James G.},
	month = jul,
	year = {2002},
	pages = {167--193},
}

@misc{lubold_identifying_2022,
	title = {Identifying the latent space geometry of network models through analysis of curvature},
	url = {http://arxiv.org/abs/2012.10559},
	doi = {10.48550/arXiv.2012.10559},
	urldate = {2025-09-02},
	publisher = {arXiv},
	author = {Lubold, Shane and Chandrasekhar, Arun G. and McCormick, Tyler H.},
	month = dec,
	year = {2022},
	note = {arXiv:2012.10559 [stat]},
}

@misc{wilkins-reeves_asymptotically_2024,
	title = {Asymptotically {Normal} {Estimation} of {Local} {Latent} {Network} {Curvature}},
	url = {http://arxiv.org/abs/2211.11673},
	doi = {10.48550/arXiv.2211.11673},
	urldate = {2025-09-02},
	publisher = {arXiv},
	author = {Wilkins-Reeves, Steven and McCormick, Tyler},
	month = aug,
	year = {2024},
	note = {arXiv:2211.11673 [stat]},
}

@article{narayan_large-scale_2011,
	title = {Large-scale curvature of networks},
	volume = {84},
	copyright = {http://link.aps.org/licenses/aps-default-license},
	issn = {1539-3755, 1550-2376},
	url = {https://link.aps.org/doi/10.1103/PhysRevE.84.066108},
	doi = {10.1103/PhysRevE.84.066108},
	language = {en},
	number = {6},
	urldate = {2025-09-02},
	journal = {Physical Review E},
	author = {Narayan, Onuttom and Saniee, Iraj},
	month = dec,
	year = {2011},
	pages = {066108},
}

@misc{kennedy_hyperbolicity_2013,
	title = {On the {Hyperbolicity} of {Large}-{Scale} {Networks}},
	url = {http://arxiv.org/abs/1307.0031},
	doi = {10.48550/arXiv.1307.0031},
	urldate = {2025-09-02},
	publisher = {arXiv},
	author = {Kennedy, W. Sean and Narayan, Onuttom and Saniee, Iraj},
	month = jun,
	year = {2013},
	note = {arXiv:1307.0031 [physics]},
}

@article{jacoby_bootstrap_2014,
	title = {Bootstrap {Confidence} {Regions} for {Multidimensional} {Scaling} {Solutions}},
	volume = {58},
	issn = {1540-5907},
	url = {https://EconPapers.repec.org/RePEc:wly:amposc:v:58:y:2014:i:1:p:264-278},
	doi = {10.1111/ajps.12056},
	number = {1},
	urldate = {2025-09-02},
	journal = {American Journal of Political Science},
	author = {Jacoby, William G. and Armstrong, David},
	year = {2014},
	note = {Publisher: John Wiley \& Sons},
	pages = {264--278},
}

@article{lusseau_bottlenose_2003,
	title = {The bottlenose dolphin community of {Doubtful} {Sound} features a large proportion of long-lasting associations - {Can} geographic isolation explain this unique trait?},
	volume = {54},
	issn = {0340-5443},
	url = {https://abdn.elsevierpure.com/en/publications/the-bottlenose-dolphin-community-of-doubtful-sound-features-a-lar},
	doi = {10.1007/s00265-003-0651-y},
	language = {English},
	urldate = {2025-09-27},
	journal = {Behavioral Ecology and Sociobiology},
	author = {Lusseau, D. and Schneider, K. and Boisseau, O. J. and Haase, P. and Slooten, E. and Dawson, S. M.},
	year = {2003},
	note = {Publisher: Springer Science and Business Media Deutschland GmbH},
	pages = {396--405},
}

@article{newman_finding_2006,
	title = {Finding community structure in networks using the eigenvectors of matrices},
	volume = {74},
	issn = {1539-3755, 1550-2376},
	url = {http://arxiv.org/abs/physics/0605087},
	doi = {10.1103/PhysRevE.74.036104},
	number = {3},
	urldate = {2025-09-27},
	journal = {Physical Review E},
	author = {Newman, M. E. J.},
	month = sep,
	year = {2006},
	note = {arXiv:physics/0605087},
	pages = {036104},
}

@article{nepusz_fuzzy_2008,
	title = {Fuzzy communities and the concept of bridgeness in complex networks},
	volume = {77},
	issn = {1539-3755, 1550-2376},
	url = {http://arxiv.org/abs/0707.1646},
	doi = {10.1103/PhysRevE.77.016107},
	number = {1},
	urldate = {2025-09-27},
	journal = {Physical Review E},
	author = {Nepusz, Tamás and Petróczi, Andrea and Négyessy, László and Bazsó, Fülöp},
	month = jan,
	year = {2008},
	note = {arXiv:0707.1646 [physics]},
	pages = {016107},
}

@article{kadushin_friendship_1995,
	title = {Friendship {Among} the {French} {Financial} {Elite}},
	volume = {60},
	issn = {0003-1224},
	url = {https://www.jstor.org/stable/2096384},
	doi = {10.2307/2096384},
	number = {2},
	urldate = {2025-09-27},
	journal = {American Sociological Review},
	author = {Kadushin, Charles},
	year = {1995},
	note = {Publisher: [American Sociological Association, Sage Publications, Inc.]},
	pages = {202--221},
}

@article{sweet_latent_2020,
	title = {A {Latent} {Space} {Network} {Model} for {Social} {Influence}},
	volume = {85},
	copyright = {https://www.cambridge.org/core/terms},
	issn = {0033-3123, 1860-0980},
	doi = {10.1007/s11336-020-09700-x},
	language = {en},
	number = {2},
	urldate = {2025-10-07},
	journal = {Psychometrika},
	author = {Sweet, Tracy and Adhikari, Samrachana},
	month = jun,
	year = {2020},
	pages = {251--274},
}

@misc{sosa_latent_2021,
	title = {A {Latent} {Space} {Model} for {Multilayer} {Network} {Data}},
	copyright = {Creative Commons Attribution 4.0 International},
	url = {https://arxiv.org/abs/2102.09560},
	doi = {10.48550/ARXIV.2102.09560},
	urldate = {2025-10-07},
	publisher = {arXiv},
	author = {Sosa, Juan and Betancourt, Brenda},
	year = {2021},
	note = {Version Number: 1},
}

@article{van_der_hoorn_ollivier-ricci_2021,
	title = {Ollivier-{Ricci} curvature convergence in random geometric graphs},
	volume = {3},
	issn = {2643-1564},
	url = {https://link.aps.org/doi/10.1103/PhysRevResearch.3.013211},
	doi = {10.1103/PhysRevResearch.3.013211},
	language = {en},
	number = {1},
	urldate = {2025-10-07},
	journal = {Physical Review Research},
	author = {Van Der Hoorn, Pim and Cunningham, William J. and Lippner, Gabor and Trugenberger, Carlo and Krioukov, Dmitri},
	month = mar,
	year = {2021},
	pages = {013211},
}

\end{document}